\documentclass[preprint,showpacs,showkeys,preprintnumbers,amsmath,amssymb]{revtex4}
 \usepackage{dcolumn}
 \usepackage{bm}

\begin{document}

 \title{ logarithmic entropy of black hole in gravity with conformal anomaly
 from quantum tunneling approach }

 \author{ Ran Li }

 \thanks{Electronic mail: liran.gm.1983@gmail.com}

 \affiliation{Department of Physics,
 Henan Normal University, Xinxiang 453007, China}

 \begin{abstract}

  Using quantum tunneling approach,
  we are able to derive the entropy with
  logarithmic term of the static spherically
  symmetric black hole in semi-classical
  Einstein equations with conformal anomaly.
  The results indicate that the logarithmic
  correction to Bekenstein-Hawking area entropy
  can be well explained by the self-gravitation.

 \end{abstract}



 \maketitle

 \section{introduction}

 Recently, a static, spherically symmetric
 black hole solution in semi-classical gravity theory with conformal
 anomaly has been presented by Cai, Cao and Ohta in \cite{CCO}.
 By studying the thermodynamics properties of the black hole solution,
 they found that the black hole entropy contains
 an usual Bekenstein-Hawking area term and a logarithmic term.
 The logarithmic correction to Bekenstein-Hawking
 area entropy has been regarded as an universal
 feature of quantum theory of gravity \cite{logarithmic}.
 They argued that such a logarithmic term might
 come from the non-local properties of effective action
 of gravity theory with conformal anomaly. This explanation
 of logarithmic term in black hole entropy is qualitative.
 Hence, this problem is worthy of a further studying.

 On the other hand, the studying of thermodynamics
 of Kehagias-Sfetsos (K-S) black hole \cite{KS} in Horava gravity
 shows that black hole entropy also has a logarithmic
 term \cite{myung-th}. By using the tunneling formulism\cite{PW}, Liu and Lu
 \cite{liulu} have been able to explain the logarithmic entropy of K-S
 black hole by the self-gravitation effect. This interpretation
 is quite different from the other approaches\cite{myung-ex,cai-ex}.
 For a recent review of the quantum tunneling method,
 one can refer to \cite{CQG}.

 In this paper, we will try to derive black hole
 entropy of the static spherically symmetric black hole
 in semi-classical Einstein gravity with conformal
 anomaly by using quantum tunneling approach.
 Firstly, we study tunneling process of massless particles
 by using null-geodesic method. By taking energy
 conservation into account, tunneling
 probability is computed in the framework of WKB approximation,
 from which black hole entropy can be read off.
 Then, by treating massive particles as
 de Broglie wave, we are able to calculate tunneling
 probability of massive particles, which is the same as that
 of massless particles. The results indicate that the logarithmic
 correction to Bekenstein-Hawking area entropy
 can be well explained by the self-gravitation.
 It implies that self-gravitation effect
 is probablely connected with the non-local property of
 quantum theory of gravity. At last, we show that
 the black hole entropy with the logarithmic term
 can also be derived by studying the fermion tunneling process
 with the back reaction effect.

 \section{black hole solutions of semi-classical Einstein equations
 with conformal anomaly}

  In this section, we will firstly give a brief review
  of static, spherically symmetric black holes in gravity with
  conformal anomaly. Cai, Cao and Ohta \cite{CCO} consider the semi-classical Einstein
  equation
  \begin{eqnarray}
   R_{\mu\nu}-\frac{1}{2}Rg_{\mu\nu}=8\pi G\langle T_{\mu\nu}\rangle\;,
  \end{eqnarray}
  where $\langle T_{\mu\nu}\rangle$ is the effective
  energy-momentum tensor by quantum loops. In four
  spacetime dimensions, the trace anomaly of energy-momentum
  tensor is given by \cite{duff,deser}
  \begin{eqnarray}
   g^{\mu\nu}\langle T_{\mu\nu}\rangle
   =\tilde{\lambda} I_{(4)}-\tilde{\alpha} E_{(4)}\;,
  \end{eqnarray}
 with $I_{(4)}=C_{\mu\nu\lambda\rho}C^{\mu\nu\lambda\rho}$
 is the type B anomaly, $E_{(4)}=R^2-4R_{\mu\nu}R^{\mu\nu}
 +R_{\mu\nu\lambda\rho}R^{\mu\nu\lambda\rho}$ is the Gauss-Bonnet
 term, which is called type A anomaly \cite{deser}.

 They obtained the static spherically symmetric black hole solution
 which is described by the metric \cite{CCO}
  \begin{eqnarray}
   ds^2=-f(r)dt^2+\frac{1}{f(r)}dr^2+r^2 d\Omega_2^2\;,
  \end{eqnarray}
  where the metric function $f(r)$ is given by
  \begin{eqnarray}
   f(r)=1-\frac{r^2}{4\alpha}
   \left(1-\sqrt{1-\frac{16\alpha GM}{r^3}+
   \frac{8\alpha Q}{r^4}}\right)\;,
  \end{eqnarray}
  with $M$ and $Q$ being the integration constants
  and $\alpha=8\pi G\tilde{\alpha}$. The parameter
  $\lambda$ has been set to zero. According to the
  detailed discussion in \cite{CCO}, the constant $M$
  is just the ADM mass of black hole and the constant $Q$
  corresponds to the $U(1)$ charge square of some
  conformal field theory.

  Under the condition of $G^2M^2>(Q-2\alpha)$, one can
  find this black hole has two horizons.
  The horizons are determined by the equation $f(r)=0$, which gives
  the locations of the inter and the outer event horizons
  as $r_{\pm}=GM\pm\sqrt{G^2M^2-(Q-2\alpha)}$.
  Then, the ADM mass $M$, Hawking temperature $T$ and entropy $S$
  of black hole can be given as \cite{CCO}
  \begin{eqnarray}
   M&=&\frac{r_+}{2G}\left(
   1+\frac{Q}{r_+^2}-\frac{\alpha}{r_+^2}\right)\;,\nonumber\\
   T&=&\frac{r_+}{4\pi(r_+^2-4\alpha)}\left(1-\frac{Q}{r_+^2}
   +\frac{2\alpha}{r_+^2}\right)\;,\nonumber\\
   S&=&\frac{A}{4G}-\frac{4\pi\alpha}{G}\ln\frac{A}{A_0}\;,
  \end{eqnarray}
  where $A=4\pi r_+^2$ is the area of the outer horizon and $A_0$
  is a constant with dimension of area. It can be easily checked that
  these thermodynamics quantities satisfy the first law of black hole
  thermodynamics $dM=TdS$.

  It is shown that the entropy is composed by
  the Bekenstein-Hawking area term and the logarithmic term.
  It is argued in \cite{CCO} that such a logarithmic term coming
  from non-local trace anomaly represents a universal feature
  of a full quantum theory of gravity.
  In the following, we will try to derive this
  logarithmic term by using the quantum tunneling approach.

 \section{logarithmic entropy from quantum tunneling}

 In this section, we will calculate the entropy of black hole
 in terms of the tunneling formalism. Three cases are considered
 respectively, the massless particles tunneling, massive particles tunneling
 and fermion tunneling.

 \subsection{massless particles case}

 To apply the null-geodesics method,
 it is necessary to choose coordinates which
 are not singular at the horizon.
 This coordinates have been
 systematically studied by Maulik K. Parikh in \cite{parikhplb}.
 Introducing the coordinate transformation
 \begin{equation}
 dt=dT-\Lambda(r)dr\;,
 \end{equation}
 where the function $\Lambda(r)$ is required to depend only on $r$
 not $t$, then, the line element (3) becomes
 \begin{eqnarray}
 ds^2&=&-f(r)dT^2+\Big(\frac{1}{f(r)}-f(r)\Lambda^2(r)\Big)dr^2\nonumber\\
 &&+2f(r)\Lambda(r)dTdr+r^2d\Omega^2\;.
 \end{eqnarray}
 Restricting the condition
 \begin{equation}
 \frac{1}{f(r)}-f(r)\Lambda^2(r)=1\;,
 \end{equation}
 one can obtain the line element in the new coordinates
 \begin{eqnarray}
 ds^2=-f(r)dT^2+2\sqrt{1-f(r)}dT dr
 +dr^2+r^2d\Omega_2^2\;.
 \end{eqnarray}

 The radial null geodesics in the new coordinates system is given by
 \begin{equation}
 \dot{r}=\frac{dr}{dT}=\pm 1-\sqrt{1-f(r)}\;,
 \end{equation}
 where the sign $+$ corresponds the outgoing null geodesics,
 while the sign $-$ corresponds the ingoing null geodesics.

 The imaginary part of the classical action
 for an outgoing positive energy particle is
 \begin{equation}
 \textrm{Im}S=\textrm{Im}\int_{r_{in}}^{r_{out}}p_r dr
 =\textrm{Im}\int_{r_{in}}^{r_{out}}\int_{0}^{p_r}dp_r'dr\;,
 \end{equation}
 where $r_{in}=GM+\sqrt{G^2M^2-(Q-2\alpha)}$ and
 $r_{out}=G(M-\omega)+\sqrt{G^2(M-\omega)^2-(Q-2\alpha)}$
 are the initial and the final radii of the black hole during the
 tunneling process. Assume that the emitted energy $\omega'\ll
 M$. According to energy conservation, the energy of background
 spacetime $M$ becomes $(M-\omega')$. Then, the outgoing radial geodesic
 is modified to be
 \begin{eqnarray}
 \dot{r}=1-\sqrt{\frac{r^2}{4\alpha}
   \left(1-\sqrt{1-\frac{16\alpha G(M-\omega')}{r^3}+
   \frac{8\alpha Q}{r^4}}\right)}\;.
 \end{eqnarray}
 From Hamilton equation
 $\dot{r}=\left.\frac{dH}{dp_r}\right|_r$, the integral can be rewritten as
 \begin{eqnarray}
 \textrm{Im}S&=&\textrm{Im}\int_{r_{in}}^{r_{out}}\int_{M}^{M-\omega}\frac{dr}{\dot{r}}dH
 \;,\nonumber\\
 &=&\textrm{Im}\int_{r_{in}}^{r_{out}}\int_{M}^{M-\omega}\frac{d(M-\omega')}{1-\sqrt{\frac{r^2}{4\alpha}
   \left(1-\sqrt{1-\frac{16\alpha G(M-\omega')}{r^3}+
   \frac{8\alpha Q}{r^4}}\right)}}dr\;,
 \end{eqnarray}
 where the energy $H$ is identified as the mass of black hole $(M-\omega')$.

 In order to perform the integral, one can introduce a new variable $u$ as
 \begin{eqnarray}
  u=\sqrt{\frac{r^2}{4\alpha}
   \left(1-\sqrt{1-\frac{16\alpha G(M-\omega')}{r^3}+
   \frac{8\alpha Q}{r^4}}\right)}\;.
 \end{eqnarray}
 Then, one can easily derive a simple relation
 \begin{eqnarray}
  d(M-\omega')=\frac{r}{G}\left(
  u-\frac{4\alpha}{r^2}u^3
  \right)du\;.
 \end{eqnarray}
 The integral becomes
 \begin{eqnarray}
  \textrm{Im}S&=&\textrm{Im}\int_{r_{in}}^{r_{out}}
  \int_{u_{in}}^{u_{out}}\frac{r}{G}\left(
  u-\frac{4\alpha}{r^2}u^3
  \right)\frac{du}{1-u} dr\nonumber\\
  &=&\textrm{Im}\int_{r_{in}}^{r_{out}}
  \int_{u_{in}}^{u_{out}}\frac{r}{G}
  \left[\frac{4\alpha}{r^2}(1+u+u^2)-1
  +\left(1-\frac{4\alpha}{r^2}\right)\frac{1}{1-u}\right]dudr
  \nonumber\\
  &=&-\frac{\pi}{G} \int_{r_{in}}^{r_{out}}\left(
  r-\frac{4\alpha}{r}\right)dr\nonumber\\
  &=&\frac{\pi}{G}\left(
  \frac{r_{in}^2-r_{out}^2}{2}-2\alpha \ln\frac{r_{in}^2}{r_{out}^2}\right)\;.
 \end{eqnarray}

 According to the WKB approximation, the tunneling probability
 for the classically forbidden trajectory is given by
 \begin{equation}
 \Gamma=\textrm{exp}(-2\textrm{Im}S)=\exp\left[\left(
  \frac{\pi}{G}r_{out}^2-\frac{\pi}{G}r_{in}^2\right)
  -\frac{4\pi\alpha}{G} \left(\ln{r_{out}^2}-\ln{r_{in}^2}\right)\right]\;.
 \end{equation}
 It is well known that the tunneling probability can also be
 expressed as the change of entropy in the tunneling process
 \begin{eqnarray}
  \Gamma=\exp(\Delta S)\;.
 \end{eqnarray}
 Then, one can read the entropy of the black hole as
 \begin{eqnarray}
  S=\frac{A}{4G}-\frac{4\pi\alpha}{G}\ln\frac{A}{A_0}\;,
 \end{eqnarray}
 which is coincide with the thermodynamic entropy given in Eq.(5).

 \subsection{massive particles case}

 To consider the quantum tunneling of massive particles,
 one should firstly deduce the corresponding radial time-like geodesic.
 According to the proposal of Zhang and Zhao in Ref.\cite{zhangzhao},
 if treating the radiative particles as de Broglie wave,
 the velocity of massive particle is given by the phase velocity,
 i.e. the radial time-like geodesic is given by
 \begin{eqnarray}
  \dot{r}=v_p=\frac{v_g}{2}=-\frac{g_{TT}}{2g_{Tr}},
 \end{eqnarray}
 where $v_g$ is the group velocity of wave package. From the line element
 (9), the radial time-like geodesic can be expressed as
 \begin{eqnarray}
  \dot{r}=\frac{1-\frac{r^2}{4\alpha}
   \left(1-\sqrt{1-\frac{16\alpha GM}{r^3}+
   \frac{8\alpha Q}{r^4}}\right)}{2\sqrt{\frac{r^2}{4\alpha}
   \left(1-\sqrt{1-\frac{16\alpha GM}{r^3}+
   \frac{8\alpha Q}{r^4}}\right)}}\;.
 \end{eqnarray}

  Then, the imaginary part of the action
  of the radiating massive particle is given by
 \begin{eqnarray}
  \textrm{Im}S&=&\textrm{Im}\int_{r_{in}}^{r_{out}}\int_{M}^{M-\omega}\frac{dr}{\dot{r}}dH
  \nonumber\\
  &=&-\textrm{Im}\int_{r_{in}}^{r_{out}}\int_{u_{in}}^{u_{out}}
  \frac{r}{G}\left(u^2-\frac{4\alpha}{r^2}u^4\right)
  \left(\frac{1}{u-1}-\frac{1}{u+1}\right)dudr
  \nonumber\\
  &=&-\frac{\pi}{G} \int_{r_{in}}^{r_{out}}\left(
  r-\frac{4\alpha}{r}\right)dr\nonumber\\
  &=&\frac{\pi}{G}\left(
  \frac{r_{in}^2-r_{out}^2}{2}-2\alpha \ln\frac{r_{in}^2}{r_{out}^2}\right)\;,
 \end{eqnarray}
 which is just the result (16) of massless particles case. So, we
 can conclude the thermodynamics entropy with the logarithmic
 term in gravity with the conformal anomaly can also be explained
 well by the self-gravitation in the semi-classical tunneling
 process of massive particles.

 \subsection{fermion case}

 This subsection is dedicated to the tunneling process
 of fermion \cite{kerner} to derive the entropy with
 logarithmic term. For simplicity,
 we consider the massless spinor field $\Psi$ obeys the
 general covariant Dirac equation \cite{ranli}
 \begin{eqnarray}
 -i\hbar\gamma^ae_a^\mu\nabla_\mu\Psi=0\;,
 \end{eqnarray}
 where $\nabla_\mu$ is the spinor covariant derivative defined by
 $\nabla_\mu=\partial_\mu+\frac{1}{4}\omega_\mu^{ab}\gamma_{[a}\gamma_{b]}$,
 and $\omega_\mu^{ab}$ is the spin connection, which can be given
 in terms of the tetrad $e_a^\mu$.
 The $\gamma$ matrices are selected as
 \begin{eqnarray*}
 \gamma^0=\left(%
 \begin{array}{cc}
  i & 0 \\
  0 & -i \\
 \end{array}%
 \right)\;,
 \gamma^1=\left(%
 \begin{array}{cc}
  0 & \sigma^3 \\
  \sigma^3 & 0 \\
 \end{array}%
 \right)\;,
 \gamma^2=\left(%
 \begin{array}{cc}
  0 & \sigma^1 \\
  \sigma^1 & 0 \\
 \end{array}%
 \right)\;,
 \gamma^3=\left(%
 \begin{array}{cc}
  0 & \sigma^2 \\
  \sigma^2 & 0 \\
 \end{array}%
 \right)\;,
 \end{eqnarray*}
 where the matrices $\sigma^k(k=1,2,3)$ are the Pauli matrices.
 According to the line element (3), the tetrad fields $e_a^\mu$ can
 be selected to be $e_{a}^{\mu}=\textrm{diag}\left(
 1/\sqrt{f(r)}\;,\;\sqrt{f(r)}\;,\;1/r\;,\;1/r\sin\theta \right)$.

 We employ the ansatz for the spin-up spinor field $\Psi$ as
 following
 \begin{eqnarray}
 \Psi=\left(A(t,r,\theta,\phi)\;,\; 0\;,\;
  B(t,r,\theta,\phi)\;,\; 0\right)^T
  \textrm{exp}\left[\frac{i}{\hbar}I(t,r,\theta,\phi)\right]\;.
 \end{eqnarray}
 In order to apply WKB approximation, we can insert the ansatz
 for spinor field $\Psi$ into the general covariant Dirac equation. Dividing by the
 exponential term and neglecting the terms with $\hbar$, one can
 arrive at the following four equations
 \begin{eqnarray}
 \left\{
  \begin{array}{ll}
  \frac{iA}{\sqrt{f(r)}}\partial_t I
  +B\sqrt{f(r)}\partial_r I =0  \;,  \\
  B\left( \partial_\theta I+\frac{i}{\sin\theta}\partial_\phi I \right)=0 \;,  \\
  A\sqrt{f(r)}\partial_r I-\frac{iB}{\sqrt{f(r)}}\partial_t I=0   \;,  \\
  A\left( \partial_\theta I+\frac{i}{\sin\theta}\partial_\phi I \right)=0  \;.
  \end{array}
 \right.
 \end{eqnarray}
 Note that although $A$ and $B$ are not constant, their
 derivatives and the components $\omega_\mu$ are all of the factor
 $\hbar$, so can be neglected to the lowest order in WKB
 approximation. The second and fourth equations indicate that
 \begin{eqnarray}
 \partial_\theta I+\frac{i}{\sin\theta}\partial_\phi I=0\;\;.
 \end{eqnarray}
 From the first and third equations one can see that
 these two equations have a non-trivial solution for $A$ and $B$ if and only if the
 determinant of the coefficient matrix vanishes. Then we can get
 \begin{eqnarray}
  \frac{(\partial_t I)^2}{f(r)}-f(r)(\partial_r I)^2=0 \;.
 \end{eqnarray}

 By separating the variables of action as
 $I=-\omega t+R(r)+\Theta(\theta,\phi)$,
 one can get
 \begin{eqnarray}
 \textrm{Im}R_\pm(r)=\pm\textrm{Im}\int\frac{\omega}{f(r)}dr=\pm\pi\frac{\omega}{f'(r_+)}\;,
 \end{eqnarray}
 where the signs $+/-$ represent the outgoing/ingoing fermions and the
 residue theorem has been used to perform the integral.
 The tunneling probability of fermions from inside to outside the
 event horizon is given by
 \begin{eqnarray}
 \Gamma=\frac{\exp\left[-2\textrm{Im}R_+\right]}
 {\exp\left[-2\textrm{Im}R_-\right]}=\exp\left[
 -\frac{4\pi\omega}{f'(r_+)}\right]\;.
 \end{eqnarray}

 From the tunneling probability, the fermionic
 spectrum of Hawking radiation
 can be deduced following the standard arguments\cite{damour,sannan},
 which results in a pure thermal spectrum.
 Then, the Hawking temperature can be determined as
 \begin{eqnarray}
 T=\frac{f'(r_+)}{4\pi}=\frac{r_+}{4\pi(r_+^2-4\alpha)}\left(1-\frac{Q}{r_+^2}
   +\frac{2\alpha}{r_+^2}\right)\;,
 \end{eqnarray}
 which is coincide with the result (5) calculated from the surface gravity.

 Now, we study the back reaction of radiating fermions to the spacetime by
 considering energy conservation. When a fermion with energy $\omega_i$
 radiates from the black hole, the mass $M$ of black hole should be
 replaced by $(M-\omega_i)$ due to energy conservation. Then the tunneling
 probability should be modified as
 \begin{eqnarray}
  \Gamma_i=\exp\left(-\frac{\omega_i}{T(M-\omega_i)}\right)\;,
 \end{eqnarray}
 where the black hole mass $M$ should be replaced by $(M-\omega_i)$ in
 the expression of Hawking temperature.
 Then, for a continuous process of lots of fermions radiating from the
 black hole, the total tunneling probability is given by
 \begin{eqnarray}
  \Gamma=\prod_i \Gamma_i&=&\exp\left(\prod_i -\frac{\omega_i}{T(M-\omega_i)}\right)
  \nonumber\\
  &=&\exp\left( -\int_{0}^{\omega}\frac{d\omega_i}{T(M-\omega_i)}\right)
  \nonumber\\
   &=&\exp\left( \int_{M}^{M-\omega}\frac{d(M-\omega_i)}{T(M-\omega_i)}\right)\;.
 \end{eqnarray}
 After performing the integral, one can obtain
 \begin{eqnarray}
 \Gamma=\exp\left[\left(
  \frac{\pi}{G}r_{out}^2-\frac{\pi}{G}r_{in}^2\right)
  -\frac{4\pi\alpha}{G} \left(\ln{r_{out}^2}-\ln{r_{in}^2}\right)\right]\;,
 \end{eqnarray}
 which is also coincide with the result (17). Then, one can conclude
 that the entropy with logarithmic term can also be
 derived by considering the fermion tunneling with back reaction.

 \section{conclusion}

 In this paper, we have studied the quantum tunneling process of the static
 spherically symmetric black hole in semi-classical
 Einstein gravity with conformal anomaly \cite{CCO}.
 This black hole exhibits a peculiar property that
 there exists a logarithmic term in its entropy.
 The tunneling probabilities for the massless and the massive particles
 are calculated by using the null-geodesic method and the phase velocity method
 respectively, from which black hole entropy can be directly read off.
 This results show that the logarithmic correction to Bekenstein-Hawking area entropy
 can be well explained by the self-gravitation in the tunneling formulism.
 It implies that self-gravitation effect is probablely connected with the non-local property of
 quantum theory of gravity. At last, we show that
 the black hole entropy with the logarithmic term
 can also be derived by studying the fermion tunneling process
 with the back reaction effect.

 \end{document}